\documentclass[
tightenlines,
preprintnumbers,
nofootinbib,
twocolumn,
 amsmath,amssymb]{revtex4-2}
\usepackage{amsmath}

\usepackage{amssymb}
\usepackage{bbold}
\usepackage{bm}
\usepackage{color,graphicx}
\usepackage{slashed}
\usepackage{comment}
\usepackage{hyperref}
\usepackage{subcaption}
\usepackage{makecell}
\usepackage[dvipsnames]{xcolor}
\usepackage{pifont}
\usepackage{caption}
\usepackage{float}
\usepackage{multirow}

\usepackage{array}
    \newcolumntype{P}[1]{>{\centering\arraybackslash}p{#1}}
    \newcolumntype{M}[1]{>{\centering\arraybackslash}m{#1}}

\DeclareMathSymbol{\mathbbE}{\mathord}{AMSb}{"45}

\DeclareMathSymbol{\mathbbH}{\mathord}{AMSb}{"48}
\DeclareMathSymbol{\mathbbR}{\mathord}{AMSb}{"52}

\begin{document}

\renewcommand\labelitemi{$\vcenter{\hbox{\tiny$\bullet$}}$}

\preprint{ANL-HEP-207858}

\title{Nuclear deep-inelastic scattering as a BSM probe via SMEFT}

\author{Supratim Das Bakshi}
\email{sdasbakshi@anl.gov}

\author{T.J. Hobbs}
\email{tim@anl.gov}

\author{Brandon Kriesten}
\email{bkriesten@anl.gov}

\affiliation{
    High Energy Physics Division, Argonne National Laboratory, Lemont, IL 60439
}

\date{\today}

\begin{abstract}
Searches for New Physics at the energy frontier have motivated theoretical development
of model-independent methods to constrain interactions beyond the Standard Model (BSM) like the Standard
Model Effective Field Theory (SMEFT), as well as associated phenomenological analyses. While these approaches are
powerful, they are also afflicted by the large space of higher-dimensional operators and the challenge
of possible degeneracies among these operators. To help ameliorate these difficulties, processes beyond
those at hadron colliders, including deep-inelastic scattering (DIS), have been identified and studied for their ability to contribute added constraining power. In this work, we extend
such investigations to charged-lepton DIS involving heavy nuclei, for which we explore the systematics of SMEFT effects for a series of representative cases. In particular, we point out that nuclear effects can reshuffle the luminosities governing partons entering SMEFT-corrected hard interactions, such that the nucleus can serve as an expanded laboratory to explore SMEFT effects complementary to free-nucleon DIS. In particular, we find that nuclear DIS possesses significant constraining power over flavor-isovector SMEFT operator combinations, with reduced nPDF dependence; similarly, improved control over nPDFs would further enhance the SMEFT reach of nuclear DIS observables. We explore projections for future (proposed) facilities like the EIC, LHeC, and FCC-eh, finding that improvements to nuclear parton distribution functions (nPDFs) can further sharpen the reach of nuclear DIS.
\end{abstract}

\maketitle

%
%
\section{Introduction}
\label{sec:intro}
The effort to test the Standard Model (SM) and identify clear signatures of New Physics beyond the SM (BSM) has engrossed the field
of particle physics and driven a wide-ranging experimental program at the Large Hadron Collider (LHC); despite tantalizing hints of anomalous physics,
however, high-energy measurements remain broadly consistent with SM baseline predictions. This dearth of unambiguous deviations from the SM has driven
an efflorescence of theoretical approaches to BSM based on parametrizations aimed at reducing the dependence on models specifying the exact dynamics
at ultraviolet scales. The most widely adopted of these approaches has been Standard Model Effective Field Theory (SMEFT), which extends the SM through the
systematic inclusion of operators of higher dimension~\cite{Buchmuller:1985jz,Grzadkowski:2010es,Brivio:2017vri,Isidori:2023pyp} beyond $d\!=\!4$ as induced by New Physics interactions at some effective scale, $\Lambda\! \gg\! m_{W,Z}$, such that the low-energy effect(s) of heavy New Physics is parametrized through these
operators and their associated Wilson coefficients.
At the High-Luminosity LHC (HL-LHC), a battery of high-energy measurements, including precise determinations of high-mass dilepton production in Drell-Yan scattering or measurement of jet-production spectra at large $p_T$ will furnish probes of SMEFT operators, such as semileptonic contact interactions~\cite{Allwicher:2022gkm,Marzocca:2026jfw}. In spite of this valuable information, the space of allowable SMEFT operators is already sizable
at the first nontrivial operator dimension ($d\!=\!6$) of general relevance for LHC measurements, which includes $\mathcal{O}(60)$ independent operators
even after flavor-symmetry assumptions. The number of relevant operators grows dramatically at higher dimensions.
Moreover, the clean interpretation of collider observables within SMEFT is further complicated by possible degeneracies among flavor and chirality structures, together with nontrivial correlations which can exist between Wilson coefficients and purely SM objects, including the parton distribution functions (PDFs) of the colliding proton~\cite{Carrazza:2019sec,Greljo:2021kvv,Iranipour:2022iak,Kassabov:2023hbm,Costantini:2024xae,Marzocca:2026jfw} or fundamental input parameters such as $\alpha_s$ or the top mass, $m_t$~\cite{Gao:2022srd}.
This has motivated machine-learning-based studies, such as foundation models, to map the growing operator degeneracies \cite{DasBakshi:2025dcg}. 

For these reasons, deep-inelastic scattering (DIS) has been highlighted as another process with sensitivity to SMEFT and the potential to augment
information from the LHC and other facilities.
In particular, DIS provides a complementary environment for probing many of the same semileptonic interactions that enter high-energy Drell-Yan production. DIS introduces dependence on Bjorken-$x$, momentum transfer $Q^2$, beam polarization, and target composition, offering handles on quark flavor and electroweak chirality that are difficult to isolate at the LHC~\cite{H1:2011ije,Boughezal:2020uwq}.
Previous works~\cite{Falkowski:2017pss,Carrazza:2019sec,Boughezal:2020uwq,Greljo:2021kvv,Boughezal:2021kla,Boughezal:2022pmb,Bissolotti:2023vdw} have examined DIS off the proton or deuteron within SMEFT and explored complementarities with hadron-collider processes, or
explored implications of HERA to constrain lepton-quark contact interactions, including in simultaneous fits with the proton PDFs~\cite{H1:2011ije,ZEUS:2019cou}.

\begin{figure}[t]
    \centering
    \includegraphics[width=\columnwidth]{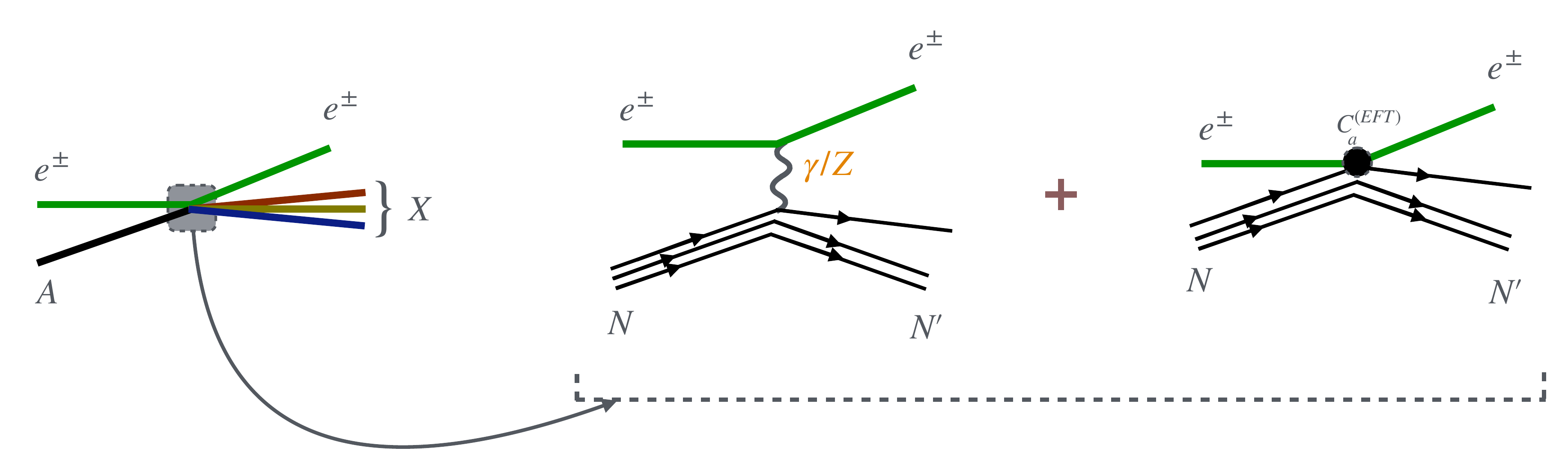}
    \caption{Schematic representation of neutral-current nuclear DIS in the presence of SMEFT corrections. The lepton scatters from a nuclear target through Standard Model $\gamma/Z$ exchange and short-distance four-fermion interactions parameterized by Wilson coefficients $C_a^{\rm EFT}$.}
    \label{fig:repr_fig0}
\end{figure}

In this
study, we investigate the unique problem of DIS from a range of heavy {\it nuclear} targets. Crucially, nuclear DIS (nDIS) entails both the challenge of nuclear-medium effects and associated uncertainties, but also introduces additional degrees-of-freedom against which the possible crosstalk of SMEFT-mediated interactions and SM effects related to partonic densities might be explored further.
The fundamental idea of this work is to treat SMEFT contributions to nDIS as arising purely through short-distance modifications of the hard lepton-quark interaction, and as thus factorizable from the long-distance physics encoded in nuclear parton distribution functions (nPDFs) as shown schematically in Fig.~\ref{fig:repr_fig0}. Since variations in both SMEFT parameters and nPDFs can induce flavor- and kinematics-dependent shifts in DIS structure functions~\cite{Kriesten:2024ist}, our central physics question is whether deviations in nDIS can be unambiguously tied to new short-distance physics, as well as when and how such effects might be hidden, mimicked, or reshuffled by nuclear structure effects.
Relatedly, we also explore to what extent nDIS supplies additional information beyond that provided by free-nucleon
DIS which might be useful to further constrain or interpret SMEFT effects.

Specifically, the nPDFs probed in nDIS are themselves $x$-, $Q^2$-, mass number $A$-, and parton flavor-dependent, and control the effective parton-level luminosities of quarks entering the short-distance interaction potentially modified by SMEFT. SMEFT-induced shifts of the hard lepton-quark interaction can therefore interweave with nuclear effects as parametrized by the nPDFs in such a way that nPDFs may either enhance or mitigate structure function-level shifts associated with SMEFT, such that nDIS can both serve as a potentially sensitive probe of BSM as well as a nontrivial testing ground for its interpretation in conjunction with other observables.
In this picture, the combined dependence of the nPDFs on $x$, $Q^2$, parton flavor ($i$), and mass number ($A$) provides an extended lever arm governing the contributions to the full nDIS structure function coming from particular SMEFT operators
linked to specific parton flavors; as we highlight below, this added leverage allows us to tune for specific flavor-dependent BSM scenarios and observe that they lead to distinctive behaviors in the nDIS structure functions.

Future lepton-hadron colliders will provide a particularly suitable environment for the observables we explore here, owing to their ability to perform precision DIS measurements with polarized lepton beams and a range of proton, deuteron, and heavy-nuclear targets~\cite{LHeC:2020van,AbdulKhalek:2021gbh}. Their broad kinematic coverage in $x$ and $Q^2$, together with target and polarization control, offers handles on flavor and electroweak chiral structure that are complementary to those available at the LHC.

The rest of our paper is organized as follows: in Section~\ref{sec:nDIS}, we outline our formalism for SMEFT-corrected
nuclear structure functions within the context of QCD factorization, including parton-level nuclear-medium effects as
parametrized by nPDFs; in Section~\ref{sec:pheno}, we present our main numerical results for the SMEFT signal in nDIS,
including the $A$-number and isospin dependence; in Section~\ref{sec:SMEFT}, we match representative UV completions
onto the relevant Wilson coefficients to explore the model-{\it dependent} implications associated with specific BSM
scenarios; and Section~\ref{sec:conc} summarizes our conclusions and identifies directions for future work.

\section{nDIS factorization in the presence of SMEFT corrections}
\label{sec:nDIS}
In this work, we focus on inclusive neutral-current deep-inelastic scattering of charged leptons from nuclei,
\begin{align}
e^\pm(k) + A(P_A) \to e^\pm(k') + X\ ,
\end{align}
which is characterized via external kinematics including the momentum transfer, $q=k-k'$ and virtuality $Q^2=-q^2$; the per-nucleon Bjorken fraction, $x=Q^2/(2P\cdot q)$, where the nucleon momentum is $P = P_{A} / A$; and the inelasticity, $y=P_A\cdot q/P_A\cdot k$. In the hard-DIS kinematical regime, $Q^2\gg \Lambda_{\rm QCD}^2$, the reduced cross section can be expressed in terms of DIS structure functions like $F_{2,A}(x,Q^2)$, which encode the partonic response of the nuclear target as given by the imaginary part of the forward Compton amplitude of the nuclear DIS process.

At leading twist, the nuclear structure function can be written as a convolution of short-distance coefficient functions with nPDFs giving the partonic densities of the corresponding nuclei. In the presence of SMEFT corrections, we organize the nuclear structure functions up to corrections of $\mathcal{O}(\Lambda^{-4})$ as
\begin{equation}
    F_{2,A}(x,Q^{2}) = F_{2,A}^{\text{SM}}(x,Q^{2}) + \Delta F_{2,A}^{(6),\rm int}(x,Q^{2})\ ,
    \label{eq:SF-general}
\end{equation}
in which
\begin{eqnarray}
    \label{eq:fact}
    \Delta F_{2,A}^{\rm (6),int}(x,Q^{2}) &=& x \sum_{i \in \mathcal{P}}\Bigg[H_{2,i}^{\rm (6),int}\Big(\xi,Q^{2};\Big\{\frac{C_{a}}{\Lambda^{2}} \Big\} \Big) \\
    &&\hspace{1.85cm}\otimes f_{i/A}\left(\frac{x}{\xi},Q^{2}\right) \Bigg]\ . \nonumber
\end{eqnarray}
Here $\mathcal{P}$ denotes the full set of partonic flavors, $\{g\} \cup \{q,\bar{q} | q \in \mathcal{Q} \}$, where $\mathcal{Q}$ is the set of active quark flavors; $f_{i/A}$ is the nPDF of parton flavor $i$, defined per nucleon in a nucleus of mass number $A$; the index $a$ labels the dimension-6 SMEFT operators with corresponding Wilson coefficients $C_{a}$;
and $H_{2,i}^{\rm (6),int}$ is the short-distance coefficient function for partonic channel $i$ at dimension-6. We note that the next-order SMEFT corrections, $\mathcal{O}(\Lambda^{-4})$ after Eq.~(\ref{eq:SF-general}), include contributions from single (linear) insertions of the dimension-8 operators as well as double insertions of the dimension-6 operators. In this work, we highlight the SMEFT-nPDF interplay at leading order, where the flavor dependence can be written most explicitly; here, $H_{2,i}^{\rm (6),int}$ reduces to an electroweak weight for each active quark flavor, $i\in \{q,\bar{q} | q\in \mathcal{Q}\}$, gluon-mediated corrections appear at higher perturbative orders in QCD, and the convolution becomes a product with $x\left[f_{q/A}+f_{\bar q/A}\right]$, as we show in more detail in Eqs.~(\ref{eq:smeftmod}) and (\ref{eq:F2SM}) below.

The essential assumption underlying this approach is that possible SMEFT effects are localized to the hard scattering, while the long-distance QCD and nuclear dynamics remain encoded in the nPDFs. This separation is justified in the DIS regime where the momentum transfer is perturbative, $Q^2\!\gg\! \Lambda_{\rm QCD}^2$, and the EFT expansion remains valid, \(Q^2\!\ll\! \Lambda^2\). Within this regime, nDIS provides a controlled setting in which short-distance SMEFT effects and nuclear-medium corrections can be compared within the same observable. 

This factorization ansatz foregrounds the conceptual core explored in this study as introduced in Sec.~\ref{sec:intro}.
A nonzero SMEFT coefficient modifies the hard interaction, $H_{2,i}^{\rm (6),int}$, while nuclear effects modify the effective quark content of the target relative to the free nucleon.
Since these contributions generate distinct scenarios for the dependence on $x$, $Q^2$, parton flavor $i$, and $A$, nDIS provides a valuable setting in which the interplay between short-distance New-Physics effects and nuclear-medium corrections can be systematically probed and potentially disentangled.

For neutral-current DIS, at leading power, the relevant SMEFT effects that are most studied arise from semileptonic four-fermion operators. Such interactions take the form
\begin{align}\label{eq:dim6op}
\mathcal{L}_{\rm SMEFT} & \supset \frac{C^{q\ell}_{\lambda\lambda'}}{\Lambda^2} \left(\bar q_\lambda \gamma_\mu q_\lambda\right) \left(\bar \ell_{\lambda'} \gamma^\mu \ell_{\lambda'}\right),
\end{align}
where $\lambda=\pm 1$ and $\lambda'=\pm 1$ label the quark and lepton chiralities, respectively, and we denote by $q$ a specific active quark flavor, in contrast to the general partonic index $i$. Here, $\lambda=\pm1 $ denotes either a right- or left-handed fermion, respectively. At leading order in the SMEFT expansion, these operators interfere with the Standard Model photon- and $Z$-exchange amplitudes and induce a correction to the (n)DIS structure functions.

For an electron beam scattering from nucleus $A$, the general expression for the correction to the $F_2$ structure function from a single operator of Eq.~(\ref{eq:dim6op}) with quark flavor $q$ and explicit helicity dependence is~\cite{Carrazza:2019sec,Greljo:2021kvv}
\begin{widetext}
\begin{equation}
\Delta F_{2,A}^{q,\lambda\lambda'}(x,Q^2)  = \frac{C^{q\ell}_{\lambda\lambda'}}{\Lambda^2} \frac{Q^2}{2e^2}  \bigg( e_q - K_Z \left(V_e-\lambda' A_e\right)  \left(V_q-\lambda A_q\right) \bigg)\ x\left[ f_{q/A}(x,Q^2)+f_{\bar q/A}(x,Q^2) \right]\ ,
\label{eq:smeftmod}
\end{equation}
\end{widetext}
where $f_{q/A}$ and $f_{\bar q/A}$ are per-nucleon nPDFs as defined in more detail in Eq.~(\ref{eq:fA}) below; in addition, \linebreak
$e^2\!=\!4\pi\alpha$ and $e_q$ is the quark electric charge in units of the positron charge. The total correction is the sum over quark flavors and chiral structures, \linebreak
$\displaystyle \Delta F_{2,A}^{\rm (6),int}(x,Q^{2}) =\sum_{q\in\mathcal{Q}} \sum_{\lambda,\lambda' = \pm 1}\Delta F_{2,A}^{q,\lambda\lambda'}(x,Q^{2})$, with one Wilson coefficient for each flavor and chirality combination, and
\begin{align}
K_Z =\frac{Q^2}{\sin^2(2\theta_W)\left(Q^2+m_Z^2\right)}\ .
\label{eq:KZ}
\end{align}
The electroweak vector and axial couplings are
\begin{align}
V_e=-\frac{1}{2}+2\sin^2\theta_W\ ,
\qquad A_e=-\frac{1}{2}\ , \nonumber
\end{align}

\begin{align}
V_q=I_3^q-2e_q\sin^2\theta_W\ ,
\qquad
A_q=I_3^q\ .
\label{eq:Vq}
\end{align}
We have verified that Eqs.~(\ref{eq:smeftmod}) and (\ref{eq:KZ}), as well as the SM structure function of Eq.~(\ref{eq:F2SM}) below, agree exactly with Refs.~\cite{Carrazza:2019sec,Boughezal:2022pmb} for an unpolarized electron beam; in Ref.~\cite{Boughezal:2022pmb} this requires averaging over the electron helicity.

For the numerical studies presented below, we adopt an electroweak scheme in which the electromagnetic coupling, $\alpha$, Fermi constant, $G_F$, and $Z$-boson mass, $m_Z$, are taken as input parameters. In this context, the Weinberg angle is written as
    $\text{sin}^2 \, 2 \theta_W = {4 \pi \alpha}/({\sqrt{2} G_F m_Z^2})$.

In the following sections, we use the above formalism to isolate effects of SMEFT corrections to the neutral-current nDIS structure functions, and compare these against effects associated with parton-level nuclear structure. We focus on the semileptonic four-fermion operators to which nDIS is especially sensitive; we do so on the logic that nDIS constrains the flavor and chirality structure of lepton-quark contact interactions in complementary fashion to high-energy processes like Drell-Yan scattering.
The nPDFs are also an essential input to the calculations below, in that they parametrize the long-distance partonic structure of nuclei necessary to compute the expressions appearing in Eq.~(\ref{eq:smeftmod}).
Modern global analyses, such as EPPS21~\cite{Eskola:2021nhw}, nCTEQ15~\cite{Kovarik:2015cma}, and nNNPDF2.0~\cite{AbdulKhalek:2020yuc}, extract these distributions from a combination of fixed-target DIS, Drell-Yan, proton-nucleus, and other collider data, with different assumptions about parametrization, flavor separation, and uncertainty propagation~\cite{Klasen:2024xqn}. These fits show that nuclear effects depend nontrivially on Bjorken-$x$, the hard scale $Q^2$, the nuclear mass number $A$, and parton flavor.

The nPDFs are typically obtained in two principal ways: either by analyzing a nuclear correction on top of a free-nucleon baseline, or by directly fitting the PDFs of an arbitrary nucleus, $A$. Correspondingly, the nuclear modifications are therefore parametrized either through a nuclear modification factor applied to the free-proton PDF~\cite{Eskola:2009uj},
\begin{align}
f_q^{p/A}(x,Q^2)=R_q^{p/A}(x,Q^2) f_{q/p}(x,Q^2)\ ,
\end{align}
\noindent or by fitting the bound-proton PDFs $f_q^{p/A}$ directly, from which the full nPDF is built as~\cite{Kovarik:2015cma,AbdulKhalek:2020yuc},
\begin{align}
f_{q/A}(x,Q^2) = \frac{Z}{A} f_q^{p/A}(x,Q^2) + \frac{A-Z}{A} f_q^{n/A}(x,Q^2)\ .
\label{eq:fA}
\end{align}
Equivalently, one may define a nuclear modification ratio by comparing the per-nucleon nPDF, $f_{q/A}$ of Eq.~(\ref{eq:fA}), to the incoherent free-nucleon baseline,
\begin{align}
R_{q/A}(x,Q^2) & = \frac{f_{q/A}(x,Q^2)} {f_{q/A}^{(0)}(x,Q^{2})}, \nonumber \\
f_{q/A}^{(0)} & = \frac{Z}{A}f_{q/p} + \frac{A-Z}{A}f_{q/n}
\label{eq:RqA}
\end{align}
where the free-neutron PDFs follow from the proton ones by isospin symmetry, $f_{u/n}=f_{d/p}$ and $f_{d/n}=f_{u/p}$. In the absence of nuclear effects, one would have $R_{q/A}=1$, while deviations from unity encode the modification of quark and antiquark distributions inside the nuclear medium. These deviations include shadowing at small $x$, antishadowing at intermediate $x$, the EMC effect at larger $x$, and Fermi-motion effects as $x\to 1$.

The flavor dependence of these nuclear modifications is particularly important for SMEFT studies. Semileptonic four-fermion operators couple differently to up-type and down-type quarks, and to different electroweak chiral structures. Therefore, any flavor-dependent change in the nPDFs can alter the apparent sensitivity to different Wilson coefficients. For example, a nuclear enhancement or suppression of the effective up-quark distribution changes the weight of operators involving electron-up-quark contact interactions, while analogous effects in the down-quark sector modify the sensitivity to electron-down-quark operators.

In practice, this motivates comparing predictions obtained with different nPDF determinations to illustrate the underlying nonperturbative model dependence and other analysis choices. For instance, nPDF analyses differ in their approach to uncertainty determination (Hessian as with EPPS and nCTEQ {\it vs.}~Monte Carlo as with nNNPDF) as well as in selections of data to drive flavor sets and internal theory choices. Future lepton-nucleus measurements at the EIC are expected to further constrain nPDFs and reduce some of this model uncertainty, including in regions of small and large $x$ and to further enhance parton-level flavor-separation~\cite{AbdulKhalek:2019mzd,AbdulKhalek:2021gbh}.

In Sec.~\ref{sec:pheno} below, we identify kinematic regions where the SMEFT correction dominates over nuclear-PDF uncertainty, as well as regions where the two effects are degenerate.

\section{Nuclear structure function effects induced by SMEFT}
\label{sec:pheno}
Having introduced the SMEFT formalism for nDIS in Sec.~\ref{sec:nDIS}, we now turn to its phenomenological implications for nuclear structure functions.

\subsection{Numerical setup}\label{sec:setup}

\paragraph{Operators and benchmarks.}
In the analysis below, we activate only right-handed current-current operators, {\it i.e.},\ $\lambda=\lambda'=+1$ in Eq.~(\ref{eq:dim6op}) and therefore drop the $\lambda=+1, \lambda'=+1$ notation and refer to $\Delta F_{2,A}^{++} = \Delta F_{2,A}$ for the remainder of the manuscript; we denote $C_{eq}\equiv C^{q\ell}_{++}$ for $\ell=e$ and $q=u,d,s,c$. As a representative subset of the full Wilson coefficient parameter space, we consider four benchmark scenarios for specific operator behaviors:
(i) $C_{ed}=+1$ ($d$-only or $d$-philic);
(ii) $C_{eu}=C_{ed}=+1$ (same-sign or isoscalar);
(iii) $C_{eu}=+1,\ C_{ed}=-1$ (isovector); and
(iv) $C_{eu}=C_{ec}=+1,\ C_{ed}=C_{es}=-1$ (isovector extended to the second generation), with $\Lambda=1$~TeV, {\it i.e.},\ $|C_{eq}|/\Lambda^2=1~{\rm TeV}^{-2}$. Unless stated otherwise, we take $Q^2\!=\!10^4$~GeV$^2$, comparable to the uppermost kinematical reach typical of the EIC. We also stress that our benchmark Wilson
coefficient values are selected to normalize the SMEFT effects explored here to illustrate the nPDF interplay, given the fact that the relevant $\Lambda/\sqrt{|C|}$ have been independently probed, {\it e.g.}, in high-mass Drell-Yan at the LHC. In fact, a crucial point of our study is the relationship: improvements in nPDF uncertainties directly relate to an extended reach in $\Lambda/\sqrt{|C|}$, with an added lever arm associated with $A$ dependence in the case of flavor-dependent nuclear corrections akin to those analyzed in light nuclei~\cite{Cocuzza:2026vey}.\\

\paragraph{SM baseline.}

We examine relative effects by normalizing to the leading-order neutral-current structure function for an unpolarized electron beam, including photon, $\gamma Z$-interference and $Z$ exchange,
\begin{align}
\label{eq:F2SM}
F_{2,A}^{\rm SM} &= \sum_{q=u,d,s,c,b} x\left[f_{q/A}+f_{\bar q/A}\right] \\
&\times \Big[e_q^2 - 2e_q V_e V_q K_Z + (V_e^2+A_e^2)(V_q^2+A_q^2)K_Z^2\Big], \nonumber
\end{align}
with $K_Z$ and the couplings of Eqs.~(\ref{eq:KZ})--(\ref{eq:Vq}). With $\alpha(m_Z)=1/127.951$, $G_F=1.1663787\times10^{-5}$~GeV$^{-2}$ and $m_Z=91.1876$~GeV we obtain $\sin^2\theta_W=0.2336$ and $K_Z=0.763$ at $Q^2=10^4$~GeV$^2$. We note that the inclusion of the $b$ quark in Eq.~(\ref{eq:F2SM}) reduces $|\Delta F_2/F_2^{\rm SM}|$ by $\approx 10\%$ at $x\sim10^{-4}$ and has a negligible effect for $x\gtrsim0.1$. In the subsequent discussion, we simplify our notation by dropping the $_{A}$ subscript and the $^{\rm (6),int}$ superscript on $\Delta F_{2}$, which henceforth represents the linear dimension-6 SMEFT correction to the $F_{2}$ structure function, and $F_{2}^{\rm SM}$ remains the baseline Standard Model calculation.\\

\paragraph{Nuclear PDFs.}
To explore the dependence on chosen nPDF, we take a representative sampling of NLO sets\footnote{We take a sufficient selection of available nPDF sets to qualitatively highlight the dependence in our calculations, aiming for a mix of sets with different uncertainty methods, data-set selections, and other choices from the past several years; at the same time, we are not exhaustive in including all available fits \cite{Buckley:2014ana}.}, including nCTEQ15~\cite{Kovarik:2015cma} as well as its corresponding 2026 release, nCTEQ26~\cite{nCTEQ:2026ewr,nCTEQ26:grids}; EPPS21~\cite{Eskola:2021nhw} (with the CT18A proton baseline~\cite{Hou:2019efy}); nNNPDF2.0~\cite{AbdulKhalek:2020yuc}; and TUJU19~\cite{Walt:2019slu}.
We note that these sets differ in their approach to implementing nuclear effects as discussed in Sec.~\ref{sec:nDIS}: the nCTEQ15 grids are bound-proton PDFs, $f_{q}^{p/A}$, from which we build the per-nucleon distribution with isospin symmetry, $f_{u/A}=\left[Z f_{u}^{p/A}+(A-Z)f_{d}^{p/A}\right]\big/A$ (and $u\leftrightarrow d$), 
whereas the nCTEQ26, EPPS21, nNNPDF2.0 and TUJU19 analyses directly give the per-nucleon nPDF of the full nucleus.
This can be confirmed explicitly by consulting the large-$x$ valence-PDF ratio, $d_v/u_v$, which is $\approx0.5$ for the nCTEQ15 grids and $>1$ for the others in $^{208}$Pb.
We stress this verification given the fact that, at large $x$, $\Delta F_2/F_2^{\rm SM}$ is especially sensitive to the nuclear $d/u$ ratio.\\

\paragraph{Uncertainties.}
The nPDF uncertainties are evaluated at 68\% CL: for the Hessian sets (nCTEQ15, nCTEQ26, EPPS21, TUJU19) we use asymmetric eigenvector errors rescaled from their native 90\% CL by a factor $1/1.645$, and for nNNPDF2.0 the standard deviation over its 250 replicas. The ratio $\Delta F_2/F_2^{\rm SM}$ is evaluated member by member, so that correlations between numerator
and denominator are retained.\\

\begin{figure*}[!tbh]
    \centering
    \includegraphics[width=\linewidth]{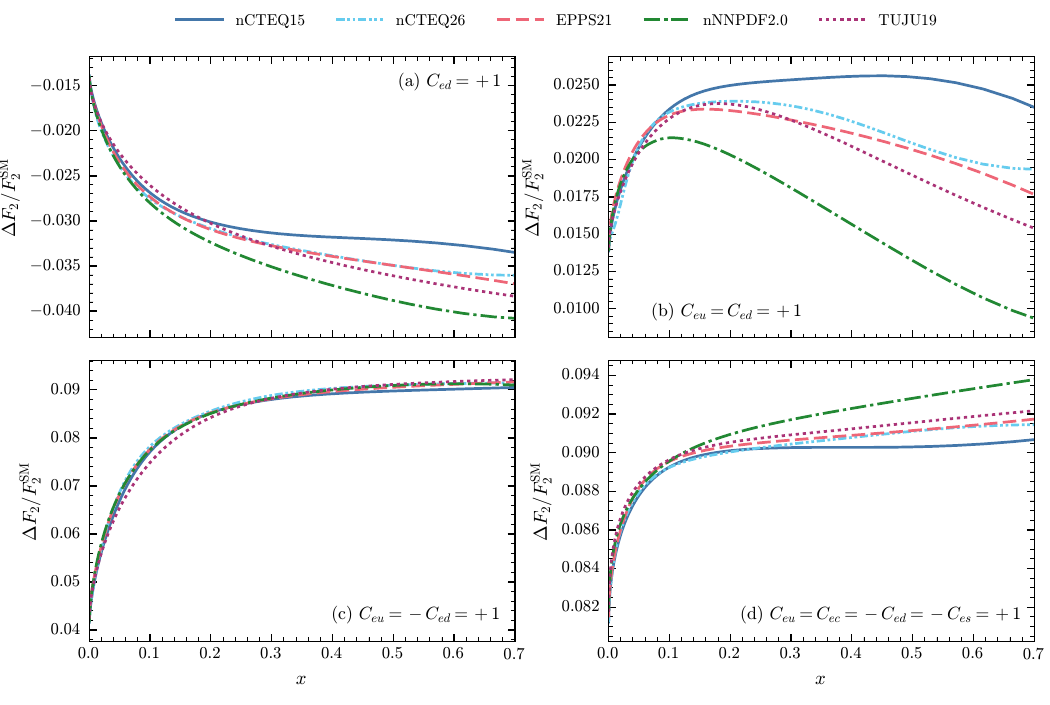}
    \caption{The SMEFT-induced fractional shift, $\Delta F_2/F_2^{\rm SM}$, for a $^{208}$Pb target at $Q^2=10^4$~GeV$^2$ and $\Lambda=1$~TeV, for four SMEFT scenarios: (a) $d$-only, $C_{ed}=+1$; (b) same-sign or isoscalar, $C_{eu}=C_{ed}=+1$, leading to partial cancellation between the up- and down-quark contributions; (c) isovector, $C_{eu}=+1$, $C_{ed}=-1$, as might be generated by a tree-level spin-1 mediator; and (d) the isovector scenario extended to the second generation, $C_{eu}=C_{ec}=+1$, $C_{ed}=C_{es}=-1$. Curves show the central members of the nCTEQ15, nCTEQ26, EPPS21, nNNPDF2.0 and TUJU19 nPDF sets; the $F_2^{\rm SM}$ baseline is the full neutral-current structure function of Eq.~(\ref{eq:F2SM}).}\label{fig:relshift_Pb208_Q10000}
\end{figure*}

\paragraph{Kinematic reach of DIS colliders.}
For the projections below, we require the kinematical parameters and reach of some representative DIS facilities.
Since $y\!=\!Q^2/(x s_{eN})\!\le\!1$, with $s_{eN}$ the electron-nucleon center-of-mass energy squared, a given $Q^2$ is accessible only for $x\ge Q^2/s_{eN}$. For a lead beam the energy per nucleon is lower than for protons, $\sqrt{s_{eN}}\simeq89$~GeV at the EIC ($18$~GeV electrons on $110$~GeV/nucleon ions~\cite{AbdulKhalek:2021gbh}), $0.74$~TeV at the LHeC ($50$~GeV on $2.76$~TeV/nucleon) and $2.2$~TeV at the FCC-eh ($60$~GeV on $19.7$~TeV/nucleon)~\cite{LHeC:2020van,FCC:2018byv}. At $Q^2=10^4$~GeV$^2$ this gives $x\gtrsim1.8\times10^{-2}$ at the LHeC and $x\gtrsim2\times10^{-3}$ at the FCC-eh, while the EIC cannot reach this $Q^2$ with lead at all: its largest $Q^2$ at $x=0.7$ is $\simeq5.5\times10^3$~GeV$^2$. As such, the small-$x$ region of the curves below is shown to illustrate the $x$ dependence of the effects and is reachable only at the highest center-of-mass energies.\\

\paragraph{Structure function and cross section.}
Experiments measure the reduced neutral-current cross section, which for $e^\pm$ beams is $\sigma_r^\pm=F_2\mp(Y_-/Y_+)\,xF_3-(y^2/Y_+)F_L$ with $Y_\pm=1\pm(1-y)^2$. We note that, at leading order, $F_L=0$, and the parity-violating structure function, $xF_3$, which arises from $\gamma Z$ interference and $Z$ exchange, receives a shift from the operators of Eq.~(\ref{eq:dim6op}); $xF_3$ depends on the valence combination $x(q-\bar q)$ rather than $x(q+\bar q)$. Moreover, its weight is $Y_-/Y_+\simeq y$ for $y\ll1$, so that $\Delta F_2/F_2^{\rm SM}$ gives the fractional shift in $\sigma_r$ for small $y$. This is in fact the relevant regime for the large-$x$ region that governs our results at high energies: at $Q^2=10^4$~GeV$^2$ and $x\ge0.1$, $Y_-/Y_+\le0.2$ at the LHeC and $\le0.02$ at the FCC-eh. At the EIC and small $x$, where $y$ approaches unity, $xF_3$ contributes at a level comparable to $F_2$. We leave the $xF_3$ contribution, the dependence on the lepton charge and beam polarization, and parity-violating asymmetries, which give strong projected EIC constraints~\cite{Bissolotti:2023vdw}, to future work.

\paragraph{Approximations.}
In this analysis, we concentrate on a first illustration of the SMEFT interplay with nuclear structure considerations; we therefore work at leading order in QCD and electroweak couplings and in the SMEFT expansion, {\it i.e.},\ we keep the interference of the dimension-6 operators with the SM, and we neglect QED radiation~\cite{Mantry:2026czd}, the renormalization-group running of the Wilson coefficients between $\Lambda$ and $Q$, and target-mass and higher-twist corrections. We explore and estimate the effects of these approximations in Sec.~\ref{sec:theory}.

\subsection{SMEFT shifts in $^{208}$Pb}\label{sec:Pb}

We start by plotting in Fig.~\ref{fig:relshift_Pb208_Q10000} the fractional shift, $\Delta F_2/F_2^{\rm SM}$ in $^{208}$Pb, for each of the four SMEFT benchmark scenarios noted at the start of Sec.~\ref{sec:setup}. The direction and size of the associated SMEFT shifts follow from the flavor-dependent electroweak factors in Eq.~(\ref{eq:F2SM}) and the analogous factors in $\Delta F_2$: {\it e.g.}, at $Q^2=10^4$~GeV$^2$, the quantity $e_q-K_Z(V_e-A_e)(V_q-A_q)$ is $\approx+0.78$ for up-type and $\approx-0.39$ for down-type quarks, such that, per unit Wilson coefficient, an up-quark operator shifts $F_2$ with approximately twice the weight, and the opposite sign, of a down-quark operator; the balance of these contributions is then governed by the nPDF weightings, term-by-term. Consequently, $C_{ed}=+1$
lowers $F_2$, while $C_{eu}=+1$ raises it.

\begin{figure*}[!tbh]
    \centering
    \includegraphics[width=\linewidth]{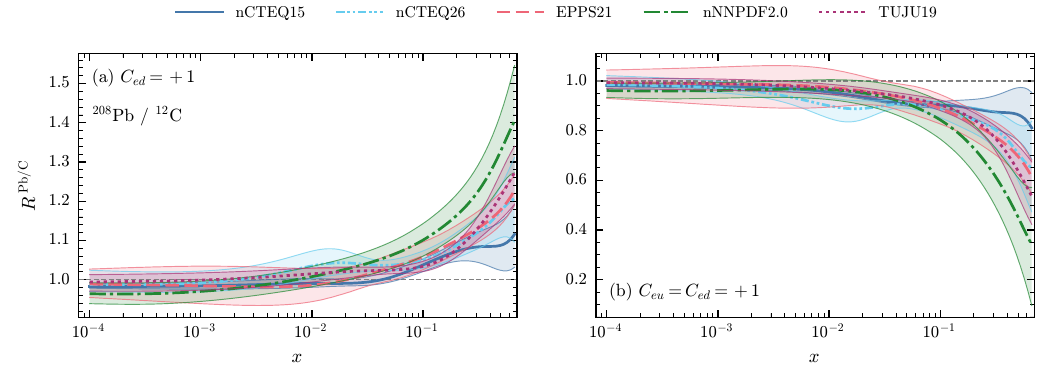}
    \caption{We plot the nuclear lever arm, $R^{\rm Pb/C\text{-}12}$, of Eq.~(\ref{eq:lever}) --- namely, the ratio of $\Delta F_2/F_2^{\rm SM}$ in $^{208}$Pb to that in $^{12}$C, versus $x$ at $Q^2=10^4$~GeV$^2$, for (a) the $d$-philic and (b) the same-sign isoscalar coupling scenarios and five nPDF sets. Bands show the 68\% CL nPDF uncertainty of $R$, evaluated member by member so that correlations between the two nuclei are retained. At linear order in $C_{eq}/\Lambda^2$, $R$ is independent of the size of the Wilson coefficients and of $\Lambda$. Its deviation from unity at large $x$ comes mainly from the neutron excess of lead; nuclear-medium effects shift it by up to $\simeq0.25$, comparable to the uncertainty of each set. The curves end at $x=0.7$, beyond which the nPDFs are less constrained.}
    \label{fig:lever_arm_PbC12_Q10000}
\end{figure*}

\paragraph{d-only or d-philic couplings.}

The SMEFT shift under this scenario is negative for all $x$, $\Delta F_2/F_2^{\rm SM}\simeq-1.4\%$ at $x=10^{-4}$, and its magnitude grows to $-3.4\%$ (nCTEQ15) or $-4.1\%$ (nNNPDF2.0) at $x\simeq0.7$.
At small $x$, the light sea is nearly flavor symmetric and the $d$-quark share of $F_2$ is fixed; at large $x$ the valence quarks dominate and the neutron excess of lead ($N/Z\simeq1.54$) raises the $d$-quark fraction, enhancing the signal.

\paragraph{Same-sign or isoscalar couplings.}

For $C_{eu}=C_{ed}=+1$ the up- and down-quark contributions have opposite signs and partially cancel, given the parton flavor-dependent electroweak weightings discussed above. The up-quark term dominates as long as $x(d+\bar d)/x(u+\bar u)<2$, so the shift is positive, $\simeq1.4\%$ at small $x$, rising to $2.2$-$2.6\%$ around $x\simeq0.1$-$0.5$. 
At larger $x$, the growing $d/u$ ratio in lead strengthens the cancellation, and the predictions separate: at $x=0.7$, they range from $0.9\%$ (nNNPDF2.0) to $2.4\%$ (nCTEQ15).

\paragraph{Opposite-sign or isovector couplings.}
For $C_{eu}=+1$, $C_{ed}=-1$, the $u$- and $d$-quark-mediated contributions add constructively, producing the largest shift(s) in the coupling scenarios we consider here; in this case, SMEFT shifts rise from $\simeq4.2\%$ at $x=10^{-4}$ to a plateau of $\simeq9\%$ for $x\gtrsim0.3$.
The rise partly reflects the decreasing weight of the $s$, $c$ and $b$ sea, which contributes to $F_2^{\rm SM}$ but not to $\Delta F_2$; at large $x$, the shift is almost independent of the nuclear $d/u$ ratio, because the SMEFT weights of $u$ and $d$ are then in nearly the same proportion as in $F_2^{\rm SM}$.
All five nPDF sets agree to within $\simeq5\%$ of the shift at every $x$.
We also stress that we illustrate the qualitative effects here for a set of natural, $\mathcal{O}(1)$,
choices of the Wilson coefficients at an arbitrary scale, $\Lambda\!=\!1$ TeV
and $Q\!=\!100$ GeV comparable to the nominal high-$Q$ reach of the EIC. Fractional SMEFT shifts can grow dramatically at still larger DIS scales in $Q$.

\paragraph{Second quark generation.}
Extending the isovector scenario to $C_{ec}=+1$, $C_{es}=-1$ adds the charm and strange sea to $\Delta F_2$, which raises the shift at small $x$ to $\simeq8\%$ and reduces the severity of the $x$ dependence as evidenced by a relatively flat $\Delta F_2 /F^\mathrm{SM}_2 \sim\!8\!-\!9.5\%$ profile over the plotted $x$ range, again with minimal dependence on the nPDF set.

\begin{figure*}[!tbh]
    \centering
    \includegraphics[width=\linewidth]{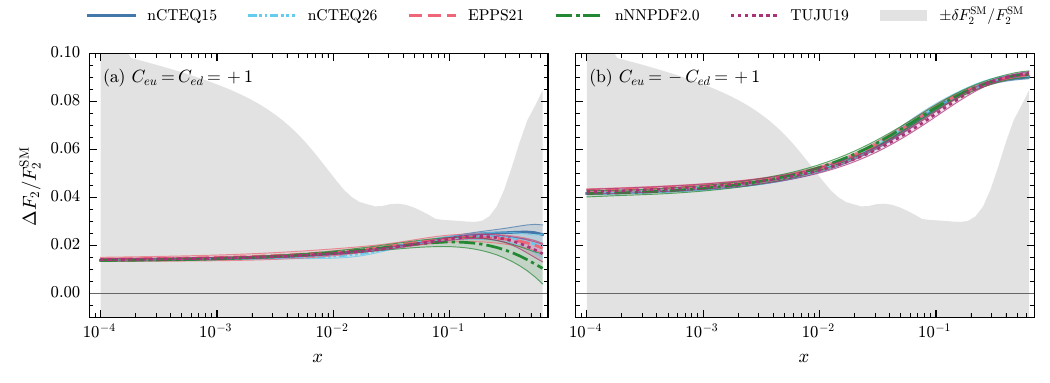}
    \caption{SMEFT-induced fractional shift $\Delta F_2/F_2^{\rm SM}$ for $^{208}$Pb at $Q^2=10^4$~GeV$^2$ and $\Lambda=1$~TeV with 68\% CL nPDF uncertainty bands of the five nPDF sets: (a) same-sign, $C_{eu}=C_{ed}=+1$; (b) isovector, $C_{eu}=+1$, $C_{ed}=-1$. The gray band is $\pm\delta F_2^{\rm SM}/F_2^{\rm SM}$, the 68\% CL nPDF uncertainty of the SM structure function (the largest among the five sets at each $x$); where a curve lies outside it, the SMEFT shift exceeds the nPDF uncertainty of the SM prediction. Both panels use the same vertical scale. Hessian uncertainties are rescaled from 90\% to 68\% CL; for nNNPDF2.0 the band is the replica standard deviation.}
    \label{fig:npdf_bands_Q10000}
\end{figure*}

Under all four SMEFT operator scenarios, the shifts scale as $C_{eq}/\Lambda^2$, so that the shifts above are reduced by a factor of four for $\Lambda=2$~TeV at a fixed choice of $C_{eq}$.
We stress that the agreement among the nPDF sets at small $x$ and the divergence at large $x$, most pronounced for the same-sign scenario, is an indication that the fitted nuclear $d/u$ ratio controls the behavior of the SMEFT-induced corrections and relates to how well a given SMEFT scenario can be separated from nuclear effects; we further quantify this with the nPDF uncertainty bands of Fig.~\ref{fig:npdf_bands_Q10000} discussed below.
We point out as well that $Q^2\!=\!10^4$~GeV$^2$ lies just beyond the reach of the EIC for a lead beam (Sec.~\ref{sec:setup}), but would be accessible at potential TeV-scale DIS machines as discussed below. At $Q^2=10^3$~GeV$^2$, which the EIC reaches for $x\gtrsim0.13$, all shifts are smaller by about a factor of ten, {\it e.g.},\ $\Delta F_2/F_2^{\rm SM}\simeq-0.3\%$ to $-0.4\%$ ($d$-only), $0.1$-$0.3\%$ (same-sign) and $0.8$-$1.0\%$ (isovector) for $0.13\le x\le0.7$ at $\Lambda=1$~TeV.

\subsection{Nuclear lever arm}\label{sec:lever}
To further isolate the nuclear dependence of the SMEFT signal, we compare a heavy, neutron-rich target with a light isoscalar one through the ratio of fractional shifts,
\begin{align}
R^{\rm Pb/C\text{-}12}(x,Q^2) = \frac{\left(\Delta F_2/F_2^{\rm SM}\right)_{^{208}{\rm Pb}}}
{\left(\Delta F_2/F_2^{\rm SM}\right)_{^{12}{\rm C}}} ,
\label{eq:lever}
\end{align}
shown in Fig.~\ref{fig:lever_arm_PbC12_Q10000} for the $d$-only and same-sign benchmarks.
We note that $R$ deviates from unity only through differences between the two nuclei.
Moreover, the factor $C_{eq}\,(Q^2/\Lambda^2)$ cancels; for a single operator, $R$ is independent of both the size and the sign of the Wilson coefficient and of $\Lambda$ (exactly, even including the squared dimension-6 term; Sec.~\ref{sec:theory}), and for two operators it depends at linear order only on their ratio, {\it e.g.},\ $C_{eu}/C_{ed}$.
The lever arm therefore does not measure the strength of New Physics but its parton-flavor dependence.

At $x\lesssim10^{-3}$, $R\simeq0.96$--$1.0$ for all sets, reflecting small differences in shadowing between lead and carbon; nCTEQ26 shows additional structure around $x\simeq0.02$, where $R$ reaches $1.04$ ($d$-only) and $0.89$ (same-sign).
For $x\gtrsim0.1$ the two benchmarks move in opposite directions: at $x=0.7$, we find $R=1.13$--$1.42$ for the $d$-only scenario and $R=0.33$--$0.79$ for the same-sign scenario.
This behavior mainly traces to the isospin structure of the associated nuclei. Carbon-12 is isoscalar, whereas lead has $N/Z\simeq1.54$, so its $d/u$ ratio at large $x$ is greater; this enhances a down-quark signal and suppresses the up-quark-dominated same-sign signal, in which the down-quark term enters with the opposite sign.
Evaluating Eq.~(\ref{eq:lever}) with free-proton PDFs combined according to $Z$ and $N$, {\it i.e.},\ with no nuclear modification at all, already gives $R\simeq1.20$--$1.27$ ($d$-only) and $R\simeq0.54$--$0.66$ (same-sign) at $x=0.7$.
Nuclear-medium effects further modify these values by up to $\simeq0.25$, with a set-dependent sign: EPPS21 lies within this isospin range, nCTEQ15 and nCTEQ26 move $R$ toward unity, and TUJU19 and nNNPDF2.0 move it further away, by up to $0.15$ and $0.23$, respectively.
These shifts are, however, comparable to the nPDF uncertainties of $R$ itself as assessed based upon the modern analyses shown here. The bands in Fig.~\ref{fig:lever_arm_PbC12_Q10000} are computed member by member in the associated error sets, so that the correlated uncertainties of lead and carbon largely cancel in the ratio. At $x=0.7$ their half-widths are $0.04$--$0.16$ ($d$-only) and $0.07$--$0.27$ (same-sign), and within each set the difference between the full and the pure-isospin value of $R$ is at most $1.3$ times the half-width. The nuclear-medium part of $R$ is therefore not resolved by any single set, while the sets differ from each other by more than their individual uncertainties, as for $\Delta F_2/F_2^{\rm SM}$ itself (Sec.~\ref{sec:bands}).

The nuclear lever arm thus separates into a part fixed by the known proton and neutron numbers, and a part tied to the flavor-dependent nuclear modifications, which is relatively weaker on the scale of current nPDF uncertainties.
The direction of the deviation from unity, $R>1$ for a down-type and $R<1$ for the up-dominated isoscalar scenario, is nevertheless robust: for $x\!\gtrsim\!0.1$, the entire 68\% CL band of every set lies on the same side of unity, so the sign of $R-1$ diagnoses the flavor structure of a signal, while its magnitude at large $x$ is limited by the less known nuclear $d/u$ ratio.
For the isovector scenario (not shown) the lever arm stays close to unity, $R=1.01$--$1.05$ for $0.3\le x\le0.7$ with uncertainties below $0.03$, because that shift is less sensitive to $d/u$ (Sec.~\ref{sec:bands}); the three operator scenarios are therefore predominantly separated by the value of $R$.

\begin{figure*}[!tbh]
    \centering
    \includegraphics[width=\linewidth]{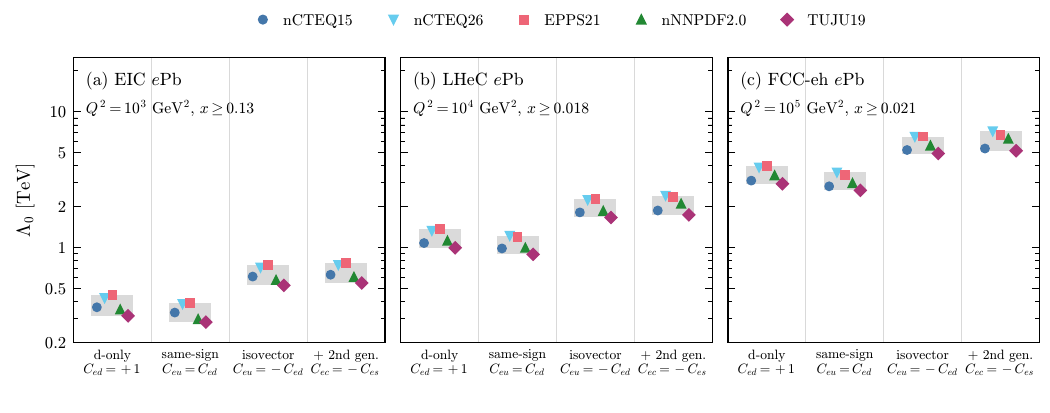}
    \caption{Reach in the New-Physics scale from $F_2$ of $^{208}$Pb: $\Lambda_0$ of Eq.~(\ref{eq:Lambda0}), the scale at which the SMEFT shift equals the 68\% CL nPDF uncertainty of $F_2^{\rm Pb}$, maximized over the $x$ range reachable with an $e$Pb beam, $Q^2/s_{eN}\le x\le0.7$, at (a) the EIC ($Q^2=10^3$~GeV$^2$), (b) the LHeC ($10^4$~GeV$^2$) and (c) the FCC-eh ($10^5$~GeV$^2$), for the $d$-only, same-sign, isovector and second-generation isovector benchmarks. Markers show the five nPDF sets and shaded bars span them. Only nPDF uncertainties are included.}
    \label{fig:lambda_reach}
\end{figure*}

\subsection{SMEFT signal versus nPDF uncertainty}\label{sec:bands}
Having outlined the calculation, quantified the SMEFT-induced structure functions shifts, and explored
the corresponding $A$ dependence, we now examine the BSM signature significance in the context of nuclear
structure uncertainties arising from the nPDFs.
Fig.~\ref{fig:npdf_bands_Q10000} shows the fractional shift in $^{208}$Pb on a common vertical scale for the same-sign (isoscalar) and isovector scenarios, with the 68\% CL uncertainty band of each of the five nPDF sets drawn around its central curve.
The gray band provides an error baseline as the relative nPDF uncertainty of the SM structure function itself, $\pm\delta F_2^{\rm SM}/F_2^{\rm SM}$, for which we take the largest of the five sets at each $x$.
The figure illustrates the separability of the SMEFT shifts (colored bands) relative to the nPDF uncertainty
in the gray band, while highlighting the fact that there is substantial dependence on the chosen scenario for
the SMEFT Wilson coefficients.

For the same-sign coupling scenario appearing in Fig.~\ref{fig:npdf_bands_Q10000}(a), the fractional SMEFT shifts predicted according to each nPDF set agree quite closely for $x\lesssim0.1$, where the shift rises from $\simeq1.4\%$ to $\simeq2.3\%$.
Above $x\simeq0.1$, however, the central curves are bracketed by a wider range and their associated uncertainty bands also widen: at $x=0.7$ the central values range from $0.9\%$ (nNNPDF2.0) to $2.4\%$ (nCTEQ15), and the nNNPDF2.0 band extends from a few tenths of a percent up to $1.7\%$.
The spread among different sets at high $x$ is therefore larger than the widest individual uncertainty, such that the choice of nPDF set is at least as consequential as the quoted uncertainty of any single set.
For the isovector scenario appearing in Fig.~\ref{fig:npdf_bands_Q10000}(b), the nPDF dependence is dramatically reduced, with all five curves lying closely together for the plotted $x$ range, rising smoothly from $\simeq4\%$ at $x=10^{-4}$ to $\simeq9\%$ at large $x$. The corresponding uncertainty bands are similarly very precise; as
such, we conclude isovector coupling scenarios enjoy much weaker nPDF dependence relative to the isoscalar, with a generally greater distinguishability from nuclear uncertainties for a given $\Lambda$.

The primary origin of the striking difference between these SMEFT scenarios is the behavior of the nuclear $d/u$ ratio.
As noted in Sec.~\ref{sec:Pb}, for equal Wilson coefficients the $u$ and $d$ quarks contribute to $\Delta F_2$ with opposite signs.
Under the same-sign scenario, the two contributions therefore partially cancel, and what remains depends strongly on the relative amount of $d$ and $u$ quarks in the nucleus: the larger the value of the nuclear $d/u$ ratio, the stronger the cancellation and the smaller the SMEFT-induced shift.
In the isovector coupling scenario, however, the two contributions add, and $u$ and $d$ quarks enter $\Delta F_2$ in nearly the same proportion as they enter $F_2^{\rm SM}$, so that a change in $d/u$ largely cancels in the ratio $\Delta F_2/F_2^{\rm SM}$.
Numerically, raising $d/u$ by $10\%$ at large $x$ lowers the same-sign shift by about $17\%$ but changes the isovector shift by less than $1\%$, {\it i.e.},\ roughly twenty times less.
The $d/u$ ratio of lead at large $x$ remains less constrained by current nPDF fits: at $x=0.7$ it ranges from $1.18\pm0.12$ (nCTEQ15) to $1.63\pm0.32$ (nNNPDF2.0). [We note that this latter per-nucleon $d/u$ ratio modestly exceeds its $N/Z\!\sim\!1.54$ upper bound, but remains consistent within uncertainties.]
This accounts for both the width of the same-sign bands and the spread between the sets in Fig.~\ref{fig:npdf_bands_Q10000}(a), where the curves are effectively ordered by the size of their nuclear $d/u$ ratio: {\it i.e.}, nCTEQ15, with the smallest $d/u$, gives the largest shift, and nNNPDF2.0, with the largest $d/u$, the smallest.

Whether a deviation can be attributed to New Physics substantially depends on how the SMEFT shifts described above compare with the nPDF uncertainty of the SM prediction --- {\it i.e.}, the gray band.
This band is widest at small $x$, where it exceeds $10\%$ at $x=10^{-4}$; it then narrows to $\simeq3\%$ for $0.05\lesssim x\lesssim0.2$, and widens again toward large $x$, reaching roughly $10\%$ near $x=0.7$.
At $\Lambda=1$~TeV, the isovector curves rise above the gray band at $x\simeq10^{-2}$ and remain above it up to $x\simeq0.6$, exceeding it by nearly a factor of three around $x\simeq0.1$--$0.3$; only at the largest $x$ does the nPDF uncertainty compete with fractional SMEFT deviations under this scenario.
The same-sign curves, in contrast, stay inside the gray band at every $x$. They come closest to its edge around $x\simeq0.1$--$0.2$, where the shift is $\simeq2.3\%$ and the band $\simeq3\%$.
The $d$-only shift of Fig.~\ref{fig:relshift_Pb208_Q10000}(a) is of similar size and likewise does not clearly exceed the gray band.

\begin{figure*}[!tbh]
    \centering
    \includegraphics[width=\linewidth]{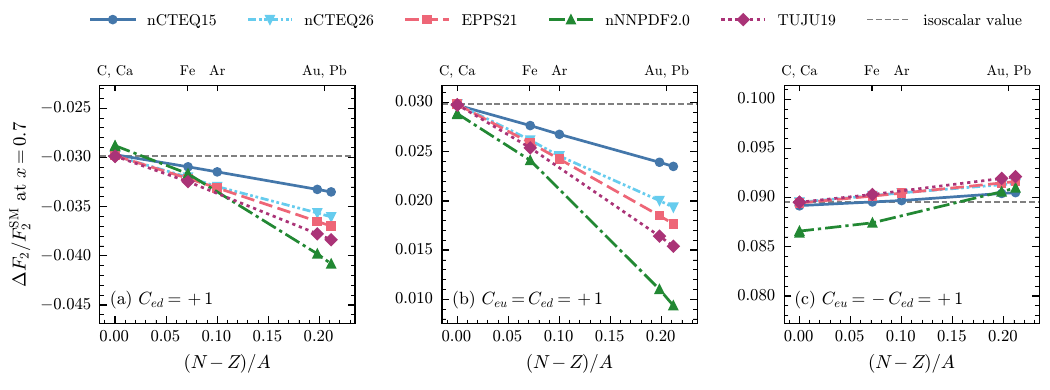}
    \caption{$\Delta F_2/F_2^{\rm SM}$ at $x=0.7$ and $Q^2=10^4$~GeV$^2$ versus the neutron excess $(N-Z)/A$, for (a) the $d$-only, (b) the same-sign and (c) the isovector benchmarks and the nuclei available in each nPDF set ($^{12}$C, $^{40}$Ca, $^{56}$Fe, $^{40}$Ar, $^{197}$Au, $^{208}$Pb; labeled on the upper axis). The dashed line is the isoscalar electroweak value $(a_u+a_d)/(s_u+s_d)$. All three panels have the same vertical span, so equal heights correspond to equal changes in $\Delta F_2/F_2^{\rm SM}$. The free proton, $(N-Z)/A=-1$, lies off scale and is discussed in the text.}
    \label{fig:A_dependence_highx_Q10000}
\end{figure*}

Such comparisons in fact allow us to define an nPDF-dependent {\it reach} with respect to the New-Physics scale, $\Lambda$. Since $\Delta F_2$ is linear in $C_{eq}/\Lambda^2$, the shift equals the 68\% CL nPDF uncertainty of $F_2$ (for a single Wilson coefficient) at
\begin{align}
\Lambda_0(x,Q^2) = 1~{\rm TeV}\times\left[\frac{\left|\Delta F_2/F_2^{\rm SM}\right|_{\Lambda=1~{\rm TeV}}}{\delta F_2/F_2}\right]^{1/2},
\label{eq:Lambda0}
\end{align}
such that contact interactions with $\Lambda<\Lambda_0$ then produce a shift larger than the nPDF uncertainty.
In practice, for a given Wilson coefficient scenario and nPDF set, we evaluate Eq.~(\ref{eq:Lambda0}) at each $x$ and take the largest value over the accessible $x$ range; this is the highest scale at which the SMEFT shift still exceeds the $68\%$ CL nPDF uncertainty of $F_2$ in at least one region of $x$.
Fig.~\ref{fig:lambda_reach} shows $\Lambda_0$ for $^{208}$Pb, maximized over the $x$ range reachable with a lead beam, at one representative $Q^2$ per collider: $10^3$~GeV$^2$ at the EIC, $10^4$~GeV$^2$ at the LHeC and $10^5$~GeV$^2$ at the FCC-eh.
Because the shift grows as $Q^2$, $\Lambda_0$ grows roughly as $Q$, from $0.3$--$0.8$~TeV at the EIC to $0.9$--$2.4$~TeV at the LHeC and $2.6$--$7.2$~TeV at the FCC-eh.
At every collider, the isovector scenario can be probed to about $1.7$ times higher scales than the $d$-only and isoscalar coupling cases, since the associated SMEFT shift is the largest and nearly insensitive to the nuclear $d/u$ ratio.
The best sensitivity comes from the window $x\simeq0.1$--$0.2$, where the nPDF uncertainty of $F_2$ is smallest, rather than from large $x$, where the shifts are larger but the nPDF uncertainty is also substantial. This observation further motivates efforts to control nPDF uncertainties at high $x$, in this region of more pronounced SMEFT shifts, as doing so has the potential to unlock the SMEFT/BSM model discrimination power of nuclear structure functions. The nPDF sets differ by up to $\simeq\!40\%$ in $\Lambda_0$, reflecting the role of nPDF modeling choices and uncertainties on the corresponding New-Physics reach.
We further note that projections based on proton-only DIS pseudodata combining multiple kinematic bins and including statistical, systematic and PDF uncertainties reach $\Lambda/\sqrt{C}\simeq2.9$~TeV ($C_{eu}$) and $1.7$~TeV ($C_{ed}$) at the EIC, using parity-violating asymmetries, and $21$ and $10$~TeV at the LHeC for single operators~\cite{Bissolotti:2023vdw}. Our $\Lambda_0$ is defined at a single $(x,Q^2)$ point and isolates the nPDF floor. It is therefore not a projected bound, but it indicates how precisely nuclear effects must be controlled before such a reach can be realized with heavy nuclei.
An absolute measurement of $F_2$ in one nucleus therefore cannot distinguish contact interactions with $\Lambda$ well above $\Lambda_0$ from a modification of the nPDFs.
This motivates ratios between nuclei (Sec.~\ref{sec:lever}), in which part of the nPDF uncertainty cancels, and, ultimately, a simultaneous fit of nPDFs and Wilson coefficients, in which New Physics could otherwise be absorbed into the nPDFs, as shown for proton PDFs in Refs.~\cite{Carrazza:2019sec,Hammou:2023heg,ZEUS:2019cou}.

\subsection{Dependence on the nucleus}\label{sec:Adep}
Fig.~\ref{fig:A_dependence_highx_Q10000} shows $\Delta F_2/F_2^{\rm SM}$ at $x=0.7$ for the nuclei provided by each nPDF set, from $^{12}$C to $^{208}$Pb, plotted against the neutron excess, $(N-Z)/A$. The values do not depend on $A$ as such but on the {\it isospin content} of the nucleus, as can be seen through several observations which we now discuss.

To start, all isoscalar nuclei give the same or very similar shifts. For $^{12}$C and $^{40}$Ca, we find $\Delta F_2/F_2^{\rm SM}=-2.9\%$ to $-3.0\%$ ($d$-only) and $+2.9\%$ to $+3.0\%$ (isoscalar coupling) for every set. This has a simple origin: in an isoscalar nucleus the per-nucleon $u$ and $d$ distributions coincide, so at large $x$, where the sea is negligible, $\Delta F_2/F_2^{\rm SM}\simeq(a_u+a_d)/(s_u+s_d)$ for the scenario in question with $a_i$ and $s_i$ representing the flavor-dependent SMEFT and SM electroweak $\big(\mathrm{factors}\big)$ in Eq.~(\ref{eq:smeftmod}) and (\ref{eq:F2SM}), respectively; $\Delta F_2/F_2^{\rm SM}$ is thus governed by a ratio of electroweak weights in which nuclear modifications cancel to the extent that they are flavor blind. For the $d$-only scenario this gives $-0.0198/0.663=-2.99\%$, in agreement with all five sets.

Moreover, for neutron-rich nuclei, the deviation from the isoscalar case grows nearly linearly with the neutron excess, $(N-Z)/A$, which is $0.07$ for $^{56}$Fe, $0.10$ for $^{40}$Ar and $0.20$--$0.21$ for $^{197}$Au and $^{208}$Pb. The two $A=40$ nuclei make this explicit: $^{40}$Ar and $^{40}$Ca have the same mass number but differ by $0.2\%$--$0.3\%$ in the $d$-only shift, and by $0.3\%$--$0.6\%$ in the same-sign shift, depending on the set. The slope, $d(\Delta F_2/F_2^{\rm SM})/d[(N-Z)/A]$, ranges from $-0.017$ (nCTEQ15) to $-0.057$ (nNNPDF2.0) for the $d$-only and from $-0.029$ to $-0.092$ for the same-sign scenario. It is nearly constant within a set for nCTEQ15, nCTEQ26 and EPPS21 but varies with $A$ for nNNPDF2.0 and TUJU19, reflecting how each set models the flavor dependence of nuclear modifications.

For the isovector scenario [shown in Fig.~\ref{fig:A_dependence_highx_Q10000}(c)], the dependence on the nucleus is naturally much weaker. The shift varies only between $8.6\%$ and $9.2\%$ from carbon to lead, and even the free proton gives $8.0\%$--$8.3\%$. Its slope in $(N-Z)/A$ is small and positive, $+0.006$ to $+0.021$, because the isovector shift is nearly insensitive to $d/u$ (Sec.~\ref{sec:bands}); with all three panels drawn on the same vertical span, these curves are nearly flat. The isoscalar value $(a_u+a_d)/(s_u+s_d)=8.96\%$ is reproduced by four of the sets, while nNNPDF2.0 lies $\simeq3\%$ below it, as it does for the $d$-only scenario, because its strange, charm and bottom distributions carry $2.4\%$ of $F_2^{\rm SM}$ at $x=0.7$, compared with at most $0.2\%$ in the other sets. An isovector signal is therefore nearly the same in every nucleus.

The $A$ dependence of the SMEFT signal at large $x$ is therefore an isospin effect. Isoscalar targets give a nearly nPDF-independent prediction fixed by the electroweak couplings, whereas neutron-rich targets probe the nuclear $d/u$ ratio, which is where the nPDF sets are more varied in their behaviors. 

\subsection{Theory uncertainties}\label{sec:theory}
The comparisons above concentrated upon the nPDF uncertainties, and we therefore estimate a series of additional theoretical uncertainties associated with the analysis choices made in our study; Table~\ref{tab:theory} summarizes the results for $^{208}$Pb and all five nPDF sets.

\paragraph{Scale variations.}
To estimate the effect of missing higher-order QCD corrections, we perform a simple 3-point variation of the factorization scale between $\mu_F=Q/2$ and $2Q$. The fractional shift $\Delta F_2/F_2^{\rm SM}$ changes by at most $1\%$ at $Q^2=10^4$~GeV$^2$ and $1.6\%$ at $10^3$~GeV$^2$, the lever arm $R^{\rm Pb/C\text{-}12}$ by less than $1\%$, and the reach $\Lambda_0$ of Fig.~\ref{fig:lambda_reach} by $1$-$4\%$. The dependence is weak because the SM and SMEFT contributions to $F_2$ are built from the same quark distributions with $Q^2$-dependent but $\mu_F$-independent weights, so that the scale dependence largely cancels in the ratio. The same holds for the quark coefficient functions at next-to-leading order, which are common to the photon, $Z$ and contact-interaction terms of $F_2$; only the gluon contribution is weighted differently in the numerator and the denominator.

\paragraph{Electroweak input scheme.}
Replacing our value $\sin^2\theta_W=0.2336$ by the $\overline{\rm MS}$ value $0.2312$ or the on-shell value $1-m_W^2/m_Z^2=0.2232$ lowers $\Delta F_2/F_2^{\rm SM}$ by $0.8$-$3.8\%$, by the same fraction for all benchmarks. For right-handed quark currents, $V_q-A_q=-2e_q\sin^2\theta_W$, so the SMEFT factor in Eq.~(\ref{eq:smeftmod}) is $e_q\left[1+2K_Z\sin^2\theta_W(V_e-A_e)\right]$, proportional to the quark charge in any scheme; the scheme therefore rescales all shifts by a common factor and cancels in $R^{\rm Pb/C\text{-}12}$ to $0.3\%$. Electroweak running of $C_{eu}$ and $C_{ed}$ between $\Lambda$ and $Q$~\cite{Jenkins:2013zja,Jenkins:2013wua,Alonso:2013hga} is of order $(\alpha/\pi)\ln(\Lambda/Q)$, at the percent level; vector currents do not run under QCD. QED radiation can change SMEFT constraints from DIS cross sections at order one, but affects spin asymmetries only at the percent level~\cite{Mantry:2026czd}; we expect our ratios to behave similarly, but have not computed it.

\paragraph{Target-mass and higher-twist corrections.}
Target-mass corrections are of order $x^2M_N^2/Q^2$, {\it i.e.},\ $4\times10^{-5}$ at $x=0.7$ and $Q^2=10^4$~GeV$^2$ and $4\times10^{-4}$ at $10^3$~GeV$^2$, and higher-twist corrections of order $1~{\rm GeV}^2/Q^2\lesssim10^{-3}$. Both are negligible at the scales considered here and largely cancel between numerator and denominator of $\Delta F_2/F_2^{\rm SM}$. The restriction to $x\le0.7$ reflects instead the limited constraints on nPDFs at larger $x$.

\paragraph{EFT truncation.}
We keep the interference of the dimension-6 operators with the SM, of order $C_{eq}/\Lambda^2$. The squared dimension-6 term adds $x(q+\bar q)\,\eta_q^2/4$ to $F_2$, with $\eta_q=C_{eq}Q^2/(e^2\Lambda^2)$, and relative to the linear term it scales as $Q^2/\Lambda^2$. At $Q^2=10^4$~GeV$^2$ and $\Lambda=1$~TeV it changes the shift by $-13\%$ ($d$-only), $+9\%$ (isovector) and $+30$-$90\%$ (same-sign), where the large value reflects the cancellation between the up- and down-quark terms at linear order, which the squared terms do not share. It is thus the largest theory uncertainty considered here: evaluated at $\Lambda=\Lambda_0$ rather than $1$~TeV, it changes the reach by $4$-$10\%$ ($d$-only), $8$-$15\%$ (same-sign) and $1$-$2\%$ (isovector). For a single operator the squared term cancels exactly in $R^{\rm Pb/C\text{-}12}$, which therefore remains independent of $\Lambda$; for two operators it does not, and for the same-sign scenario at $\Lambda=1$~TeV it changes $R$ by up to $50\%$ at $x=0.7$, decreasing as $1/\Lambda^2$. Dimension-8 contributions enter at the same order in $Q^2/\Lambda^2$ and are not included.

In summary, the scale and scheme uncertainties are generally mild, well below the nPDF uncertainty of the $d$-only and isoscalar coupling shifts and comparable to it only for the isovector scenario at large $x$, whose nPDF uncertainty in the SMEFT-shift ratio is itself at the percent level. The dominant theoretical limitation is the EFT truncation, which matters at $\Lambda\simeq1$~TeV and at the highest $Q^2$, and which is why our results at $\Lambda=1$~TeV should be read as a normalization of the linear effect.

\begin{table}[t]
    \centering
    \renewcommand{\arraystretch}{1.3}
    \begin{ruledtabular}
    \begin{tabular}{lccc}
     & $\Delta F_2/F_2^{\rm SM}$ & $R^{\rm Pb/C\text{-}12}$ & $\Lambda_0$ \\
    \hline
    Scale choice & $\le1$-$1.6\%$ & $<1\%$ & $1$-$4\%$ \\
    EW scheme & $0.8$-$3.8\%$ & $\le0.3\%$ & $\le3\%$ \\
    TMC, higher twist & $\lesssim10^{-3}$ & $\lesssim10^{-3}$ & - \\
    dim-6 squared, $\Lambda=1$~TeV & $9$-$90\%$ & $0$-$50\%$ & - \\
    dim-6 squared, $\Lambda=\Lambda_0$ & - & - & $1$-$15\%$ \\
    \hline
    nPDF (68\% CL) & $0.4$-$80\%$ & $\simeq1$-$40\%$ & - \\
    \end{tabular}
    \end{ruledtabular}
    \caption{Theory uncertainties of the fractional shift, the lever arm and the reach in $^{208}$Pb, as relative changes (ranges over benchmarks, $x\le0.7$, $Q^2=10^3$-$10^5$~GeV$^2$ and the five nPDF sets), compared with the 68\% CL nPDF uncertainty. The EFT-truncation entries at $\Lambda=1$~TeV refer to $Q^2=10^4$~GeV$^2$ and scale as $Q^2/\Lambda^2$.}
    \label{tab:theory}
\end{table}

\section{SMEFT contributions from UV models}\label{sec:SMEFT}
It is possible to relate the SMEFT scenarios explored above to concrete UV-complete theories to further drive
home the fact that nuclear DIS implicates particular SM extensions with some model-dependent discriminating power.
The pattern of SMEFT Wilson coefficients generated at low energy is dictated by the quantum numbers of whatever New-Physics state is integrated away. Since the DIS structure function $F_2$ receives corrections from the semileptonic dimension-6 operators of Eq.~(\ref{eq:dim6op}), asking which UV completions can populate a given $(C_{eu}, \, C_{ed})$ scenario is equivalent to asking which nuclear-DIS signatures are actually reachable by physical models, as opposed to being an artifact of scanning the Wilson-coefficient space freely. We consider two classes of tree-level mediators: scalar color-triplet leptoquarks, which couple to a single lepton–quark bilinear, and a color-singlet vector state, which couples independently to each SM fermion current. We comment on vector leptoquarks, whose exchange flips the sign of the generated coefficients, at the end of this section.\\[.1cm]

\begin{table*}[t]
    \renewcommand{\arraystretch}{2.0}
    \centering
    \begin{tabular}{|c|c|c|c|c|c|}
    \hline
    Heavy field &  $C_{eu} > 0 , \,  $  & $C_{eu} = 0, \,  $ & $C_{eu} > 0 , \,  $ & $C_{eu} \gtrless 0 , \,  $ & RGE generated  \\
    quantum nos.& $C_{ed} = 0$ & $C_{ed} > 0 $ & $C_{ed} > 0 $ & $C_{ed} \gtrless 0 $ & $C_{eu}$ and $C_{ed}$\\ 
    \hline
    $(3,1,-\frac{1}{3})_0 $ &  \checkmark &&&&\\
    $(3,1,-\frac{4}{3})_0$ &  & \checkmark&&&\\
    $(3,1,-\frac{1}{3})_0 \oplus (3,1,-\frac{4}{3})_0 $ &   &  & \checkmark & & \\
    $(1,1,0)_1$ &  &  &  & \checkmark &\\
    $(1,2,\frac{1}{2})_0$ & &  &  &  & \checkmark \\
    \hline    
    \end{tabular}
    \caption{BSM representations, labeled by $[SU(3)_C, SU(2)_L, U(1)_Y]_\mathrm{Spin}$, that generate the right-handed current-current operators $C_{eu}$ and $C_{ed}$, and the sign pattern that results. The first four rows generate them at tree level \cite{deBlas:2017xtg,DasBakshi:2018vni,Anisha:2021hgc}; the second Higgs doublet (last row) contributes only through renormalization-group running, with a sign that depends on its couplings and is not determined here \cite{DasBakshi:2024krs}. Vector leptoquarks, not listed, generate coefficients of negative sign (see text).}
    \label{tab:bsm1}
\end{table*}

\paragraph{Leptoquark completions \\[.5cm]}

Table~\ref{tab:bsm1} summarizes, for each UV completion considered in this work, whether $C_{eu}$ and $C_{ed}$ are generated at tree level or only through renormalization-group running, and which sign pattern results. The scalar leptoquarks realize the $d$-only, $u$-only and same-sign benchmarks with positive coefficients, the color-singlet vector allows either sign for each coefficient, and the second Higgs doublet contributes only radiatively. 

As a worked example, we consider the scalar leptoquark, $\omega_1 := (3,1,-1/3)_0$, assuming it couples only to the right-handed up quark and electron \cite{deBlas:2017xtg},
\begin{align}\label{eq:lag1}
    \mathcal{L}_{\omega_1} & \supset y^{(eu)}_{\omega_1; i,j} \omega_1^\dagger \bar{e}^C_i  u_j
\end{align}
In this case, the matching relation~\cite{deBlas:2017xtg} between the Wilson coefficient, $C_{eu}$, and the model-specific BSM coupling above is,\footnote{We use the convention $\widehat{C}=C/\Lambda^{2}$ to simplify notation.}
\begin{align}
  \widehat{C}_{\underset{ijkl}{eu}} &= \frac{(y^{(eu)}_{\omega_1; i,k})^* (y^{(eu)}_{\omega_1; j,l})}{2 M^2_{\omega_1}} \, ; \ \  \widehat{C}_{\underset{ijkl}{ed}} = 0\ .
  \label{eq:mat1}
\end{align}
\noindent 
An analogous scalar leptoquark, {$\omega_4 := (3,1,-4/3)_0$, in the notation of Ref.~\cite{deBlas:2017xtg}}, couples only to the right-handed down quark through a Yukawa interaction of the same form as Eq.~(\ref{eq:lag1}), and generates $C_{ed}$ via the same matching structure as Eq.~(\ref{eq:mat1}) (with $u \to d$, $M_{\omega_1}\to M_{{\omega_4}}$), and then leaving $C_{eu}=0$. Because $\omega_1$ and {$\omega_4$} each individually generate only a non-negative Wilson coefficient, their simultaneous exchange populates the isoscalar or same-sign coupling scenario, $C_{eu} = C_{ed} > 0$.

Because $C^{eu}_{ijkl}$ is built from a single Yukawa matrix contracted with itself, it is manifestly non-negative for the flavor-diagonal couplings used throughout this work, $C^{eu}_{iijj} \geq   0 $. The same argument applies to every scalar leptoquark: at tree level, scalar leptoquark exchange can only generate non-negative $C_{eu}$ and $C_{ed}$. One-loop matching corrections~\cite{Gherardi:2020det} are suppressed by a loop factor and do not change this conclusion for perturbative couplings.\\[.1cm]

\paragraph{Vector mediator completion \\[.3cm] }

The vector state $\mathcal{B} := (1,1,0)_1$ instead couples independently to every SM fermion current,
\begin{align}
    \mathcal{L}_{\mathcal{B}} & \supset g^{l}_{\mathcal{B}; i,j} \mathcal{B}^\mu \bar{l}_i \gamma_\mu l_j + g^{q}_{\mathcal{B}; i,j} \mathcal{B}^\mu \bar{q}_i \gamma_\mu q_j \nonumber \\
    &+ g^{e}_{\mathcal{B}; i,j} \mathcal{B}^\mu \bar{e}_i \gamma_\mu e_j + g^{d}_{\mathcal{B}; i,j} \mathcal{B}^\mu \bar{d}_i \gamma_\mu d_j \nonumber \\
    &+ g^{u}_{\mathcal{B}; i,j} \mathcal{B}^\mu \bar{u}_i \gamma_\mu u_j\ .
\end{align}
For such UV theories, the matching relation between the Wilson coefficient $C_{eu}$ and the $\mathcal{B}$ BSM coupling is~\cite{deBlas:2017xtg},
\begin{align}
  \widehat{C}_{\underset{ijkl}{eu}} &= -  \frac{(g^{u}_{\mathcal{B}; k,l}) (g^{e}_{\mathcal{B}; i,j})}{M^2_{\mathcal{B}}} \, ;  \ \ \widehat{C}_{\underset{ijkl}{ed}} &= -  \frac{(g^{d}_{\mathcal{B}; k,l}) (g^{e}_{\mathcal{B}; i,j})}{M^2_{\mathcal{B}}}\ .
\end{align}
Unlike the leptoquark case, $g^u_\mathcal{B}$, $g^d_\mathcal{B}$, and $g^e_\mathcal{B}$ are independent parameters of the same Lagrangian, so $C_{eu}$ and $C_{ed}$ carry no fixed relative sign: the sign of each is set separately by {${\rm sgn}(g^u_\mathcal{B} g^e_\mathcal{B})$ and ${\rm sgn}(g^d_\mathcal{B} g^e_\mathcal{B})$}.\\[.6cm]

\paragraph{Sign structure and which benchmarks are reachable \\[.3cm]}

For leptoquarks, the sign of the generated coefficient is fixed by the spin of the mediator. Scalar leptoquarks give $C_{eu},C_{ed}\geq0$, as shown above, whereas the exchange of a vector leptoquark coupling to a right-handed lepton-quark current, $\mathcal{U}_2=(3,1,2/3)_1$ ($\bar e\gamma_\mu d$) or $\mathcal{U}_5=(3,1,5/3)_1$ ($\bar e\gamma_\mu u$) in the notation of Ref.~\cite{deBlas:2017xtg}, gives $C_{ed}\leq0$ or $C_{eu}\leq0$, respectively, with magnitude $|g|^2/M^2$. An isovector scenario, $C_{eu}>0$ and $C_{ed}<0$, therefore cannot be generated by scalar leptoquarks alone. It requires at least one vector mediator: either the color-singlet $\mathcal{B}$ with $g^u_\mathcal{B}$ and $g^d_\mathcal{B}$ of opposite sign, or a combination of a scalar and a vector leptoquark such as $\omega_1\oplus\mathcal{U}_2$. In terms of the benchmarks of Sec.~\ref{sec:pheno}, the $d$-only and same-sign scenarios of Figs.~\ref{fig:relshift_Pb208_Q10000}-\ref{fig:A_dependence_highx_Q10000}, as well as a $u$-only scenario, are reachable by one or two scalar leptoquarks, while the isovector scenarios probe completions that contain a vector state. Combined with the results of Sec.~\ref{sec:bands}, the scenario that is most robust against nPDF uncertainty is thus also the one that points to a spin-1 mediator.\\[.1cm]

\paragraph{RGE-generated contributions: 2HDM \\[.3cm]}
Unlike the leptoquark and vector completions above, a heavy second Higgs doublet, $\varphi=(1,2,1/2)_0$, in accordance with the two-Higgs doublet model (2HDM), does not generate $C_{eu}$ or $C_{ed}$ at tree level. Integrating it out produces four-fermion operators, namely the scalar operators $\mathcal{O}_{ledq}$ and $\mathcal{O}_{lequ}^{(1)}$ and the current-current operators $\mathcal{O}_{le}$, $\mathcal{O}_{qu}^{(1,8)}$ and $\mathcal{O}_{qd}^{(1,8)}$, together with Higgs-sector operators~\cite{deBlas:2017xtg}, none of which contributes to $F_2$ at the order considered here. These operators mix into $C_{eu}$ and $C_{ed}$ under one-loop renormalization-group running from the matching scale $\Lambda$ down to the experimental scale $Q$, via the anomalous-dimension matrix computed in Refs.~\cite{Jenkins:2013zja,Jenkins:2013wua,Alonso:2013hga}. The resulting coefficients are suppressed by a loop factor and carry no fixed sign.

This provides an example of a UV completion that reaches the $(C_{eu}, C_{ed})$ scenario only radiatively. We include it in Table~\ref{tab:bsm1} to indicate this class of completions, without a sign pattern: a quantitative estimate requires the one-loop matching of the 2HDM onto the SMEFT, studied in Ref.~\cite{DasBakshi:2024krs}, combined with the running above, and is left for future work.

\section{Conclusions and Outlook}\label{sec:conc}

In this work, we have examined how SMEFT corrections to neutral-current structure functions are reshaped when the target is a nucleus rather than a free nucleon, treating the New-Physics contribution as a short-distance modification of the hard lepton-quark vertex while retaining nuclear-medium effects in the nPDFs. Because both ingredients have the potential to generate $x$-, $Q^2$-, flavor-, and $A$-dependent shifts in $F_2$, the two are not automatically separable, and we have performed a detailed investigation of when these can be disentangled as well as the necessary conditions and prospects at future facilities.

For $^{208}$Pb at $Q^2=10^4$~GeV$^2$ and $\Lambda=1$~TeV, right-handed contact interactions shift $F_2$ by $1$--$4\%$ for $d$-only and same-sign scenarios and by $4$--$9\%$ for isovector scenarios; for illustration, these behaviors assume the case of $C_i=1$ for all nonzero Wilson coefficients, such that the shifts induced by SMEFT would vary predictably for different choices of $C_i$ and $\Lambda$. Given their isoscalar and isovector nature, these SMEFT deviations are tied to the behavior of the nuclear $d/u$ ratio at large $x$. In particular, because the up- and down-quark SMEFT weights partially cancel for same-sign coefficients but enter in similar proportion as in $F_2^{\rm SM}$ for isovector coefficients, the same-sign shift is about twenty times more sensitive to $d/u$. At the same time, its nPDF uncertainty reaches $80\%$ at $x=0.7$, and the five nPDF sets disagree with each other beyond their nominal uncertainties, while the isovector shift is predicted to within a few percent by all of them.

We further quantify the significance of the New-Physics shift relative to the nPDF uncertainty of the SM structure function, which we find to be $1.6$--$13\%$ in lead.
In a single nucleus, the SMEFT shift exceeds this uncertainty only for $\Lambda$ below $0.3$-$0.8$~TeV at the EIC ($Q^2=10^3$~GeV$^2$), $0.9$--$2.4$~TeV at the LHeC ($10^4$~GeV$^2$) and $2.6$--$7.2$~TeV at the FCC-eh ($10^5$~GeV$^2$), with the isovector scenarios reaching to the highest New-Physics scales.
For a lead beam, $Q^2=10^4$~GeV$^2$ is accessible at the LHeC ($x\gtrsim1.8\times10^{-2}$) and the FCC-eh ($x\gtrsim2\times10^{-3}$); at the EIC, which reaches $Q^2\simeq10^3$~GeV$^2$ for $x\gtrsim0.13$, the shifts are about ten times smaller.

We also investigated the leverage furnished by exploring the $A$ dependence of the SMEFT shifts over nuclei of widely separated atomic number through a nuclear `lever-arm' ratio, $R^{\rm Pb/C\text{-}12}$. This ratio does not depend exactly on the size of the Wilson coefficients or $\Lambda$ at linear order (for a single operator), and its deviation from unity is governed foremost by the neutron excess of lead with a significant secondary contribution from nuclear-medium effects, which can modify the ratio by $\simeq\!0.25$, comparable to current nPDF uncertainties. Its direction, $R>1$ for down-type and $R<1$ for up-dominated scenarios, holds within the 68\% CL uncertainty of every nPDF set for $x\gtrsim0.1$. Isoscalar nuclei give a nearly nPDF-independent shift fixed by the electroweak couplings, and the dependence on the nucleus follows $(N-Z)/A$, as the two $A=40$ nuclei, $^{40}$Ar and $^{40}$Ca, make explicit.
From this behavior, we conclude that clearer flavor-dependent nuclear corrections --- as would be evidenced, {\it e.g.}, by reduced nPDF uncertainty and spread in the high-$x$ region of Fig.~\ref{fig:lever_arm_PbC12_Q10000} --- would supply stronger $A$-dependent control over the interplay of semileptonic SMEFT effects and the partonic structure of nuclei.
This latter information would also add an informative complement to SMEFT constraints from DIS or Drell-Yan from the free nucleon.

The flavor-dependent Wilson coefficient benchmarks we explored have implications for well-motivated New-Physics models, such that nDIS can furnish independent information to test these scenarios.
This can be seen from the fact that the SMEFT Wilson coefficient scenarios map onto specific classes of UV completions. Scalar leptoquarks generate only non-negative $C_{eu}$ and $C_{ed}$ and realize the $d$-only, $u$-only and same-sign (isoscalar) scenarios; vector leptoquarks generate non-positive ones; and the isovector scenario requires at least one spin-1 mediator, either a color singlet with opposite couplings to up and down quarks or a combination of scalar and vector leptoquarks. The scenario that is most stable against nuclear uncertainties is therefore that which implicates a vector mediator.

Lastly, we emphasize that this work highlights the potential New-Physics reach furnished by improved knowledge of nPDFs and their crosstalk with BSM parametrizations like SMEFT. Since the nPDF uncertainty of $F_2^{\rm SM}$ can rival the BSM signal for the benchmark scenarios we computed, a joint determination of SMEFT coefficients and nPDFs, in which New Physics could otherwise be absorbed into the nPDFs, is a natural next step, together with projected EIC, LHeC and FCC-eh uncertainties, higher-order corrections, and the extension to polarized beams and to the parity-violating structure function $xF_3$, which would add further flavor and chirality lever arms.
Such future efforts offer the prospect of using the nucleus as an expanded laboratory to further test or refine New-Physics searches and signatures related to SMEFT, beyond the free nucleon. While nuclear scattering sources unique uncertainties as we have shown, it is a double-edged sword, carrying the potential for $A$-dependent reweightings of SMEFT currents allowing these effects to be quantified and controlled still more comprehensively in the HL-LHC era and beyond.

\begin{acknowledgements}
We thank Michael Klasen for his suggestion to consider the recently released nCTEQ26 nPDF set in this study. Work at Argonne National Laboratory
was supported by the U.S.~Department of Energy under contract DE-AC02-06CH11357.
\end{acknowledgements}

%
%

\bibliographystyle{utphys}
\bibliography{references}

\end{document}